\documentclass{iaus}
\usepackage{graphicx}

\title[Galaxies at High Redshift] 
{Luminosities, Masses and Star Formation Rates of Galaxies at High Redshift}

\author[Andrew Bunker]   
{Andrew J.\ Bunker$^1$}

\affiliation{$^1$Department of Physics, University of Oxford\\
Denys Wilkinson Building, Keble Road, Oxford OX1\,3RH, U.K. \\email: {\tt a.bunker1@physics.ox.ac.uk}}

\pubyear{2012}
\volume{279}  
\pagerange{xxx--yyy}
\jname{Death of Massive Stars: Supernovae and Gamma-Ray Bursts}
\editors{P. Roming, N. Kawai \& E. Pian, eds.}
\begin{document}

\maketitle

\begin{abstract}
There has been great progress in recent years in discovering star forming galaxies
at high redshifts ($z>5$), close to the epoch of reionization of the intergalactic medium (IGM).
The WFC3 and ACS cameras on the Hubble Space Telescope have enabled Lyman
break galaxies to be robustly identified, but the UV luminosity function and star formation
rate density of this population at $z=6-8$ seems to be much lower than at $z=2-4$.
High escape fractions and a large contribution from faint galaxies below our current
detection limits would be required for star-forming galaxies to reionize the Universe.
We have also found that these galaxies have blue rest-frame UV colours, which might
indicate lower dust extinction at $z>5$. There has been some spectroscopic confirmation
of these Lyman break galaxies through Lyman-$\alpha$ emission, but the fraction
of galaxies where we see this line drops at $z>7$, perhaps due to the onset of
the Gunn-Peterson effect (where the IGM is opaque to Lyman-$\alpha$).
\end{abstract}

\firstsection 
\section{Introduction}

In the past decade, the quest to observe the most distant galaxies in the
Universe has rapidly expanded to the point where the discovery of $z\simeq
6$ star-forming galaxies has now become routine.  Deep imaging surveys
with the {\em Hubble Space Telescope (HST)} and large ground based
telescopes have revealed hundreds of galaxies at $z\simeq 6$ (Bunker et
al. 2004,  Bouwens et al. 2006, 2007). These searches typically rely on the Lyman break galaxy (LBG) technique
pioneered by Steidel and collaborators to identify star-forming galaxies
at $z\approx 3-4$ (Steidel et al. 1996, 1999). Narrow-band searches
 have also proved successful in isolating the Lyman-$\alpha$ emission line
 in redshift slices between $z=3$ and $z=7$ (e.g., Ouchi et al.\ 2008; Ota et al.\ 2008). Rapid follow-up of Gamma Ray Bursts has
 also detected objects at $z\sim 6$, and most recently one spectroscopically confirmed at $z=8.2$ (Tanvir et al.\ 2009).

 Parallel to these developments in identifying high redshift objects has been
 the discovery of the onset of the Gunn-Peterson (1965) effect. This is the near-total absorption of the
 continuum flux shortward of Lyman-$\alpha$ in sources at $z> 6.3$ due to the intergalactic
 medium (IGM) having a much larger neutral fraction at high redshift. The Gunn-Peterson trough was
  first discovered in the spectra of SDSS quasars (Becker et al.\ 2001, Fan et al.\ 2001, 2006). This defines the end of the reionization epoch, when the Universe
 transitioned from a neutral IGM. Latest results from WMAP indicate the mid-point of reionization may have occurred at $z\approx 11$ (Dunkley et al.\ 2009). The source
of necessary ionizing photons remains an open question: the number density
of high redshift quasars is insufficient at $z>6$ to achieve this (Fan et al.\ 2001, Dijkstra et al.\ 2004). Star-forming galaxies at high redshift are another potential driver of reionization, but we must
 first determine their rest-frame UV luminosity density to assess whether they are plausible sources;
 the escape fraction of ionizing photons from these galaxies, along with the slope of their UV spectra, are
 other important and poorly-constrained factors in determining whether star formation is responsible for the ionization of the IGM at high redshift.
 
 \section{The Star Formation Rate at High Redshift}

Early results on the star formation rate density at $z\approx 6$ were conflicting,
with some groups claiming little to no evolution to $z\sim 3$ (Giavalisco et al.\ 2004, Bouwens
et al.\ 2003) while other work suggested that the star formation rate density at
$z\approx 6$ was significantly lower than in the well-studied LBGs at $z=3-4$ (Stanway et al.\ 2003). The consensus which has now emerged from later studies is that the abundance of {\it
luminous} galaxies is substantially {\it less} at $z\approx 6$ than at
$z\approx 3$ (Stanway et al. 2003, Bunker et al. 2004, Bouwens et al.
2006, Yoshida et al. 2006, McLure et al. 2009).  If this trend continues
to fainter systems and higher redshifts, then it may prove challenging 
for star-forming galaxies to provide the UV flux needed to fulfill
reionisation of the intergalactic medium at (e.g.\ Bunker et al. 2004). Importantly, analysis
of the faint-end of the luminosity function at high redshift has revealed that feeble
galaxies contribute an increased fraction of the total UV luminosity, suggesting that the bulk of star formation (and hence reionizing
photons) 
likely come from lower luminosity galaxies not yet adequately
probed even in deep surveys.

\begin{figure}[b]
\includegraphics[width=2.5in]{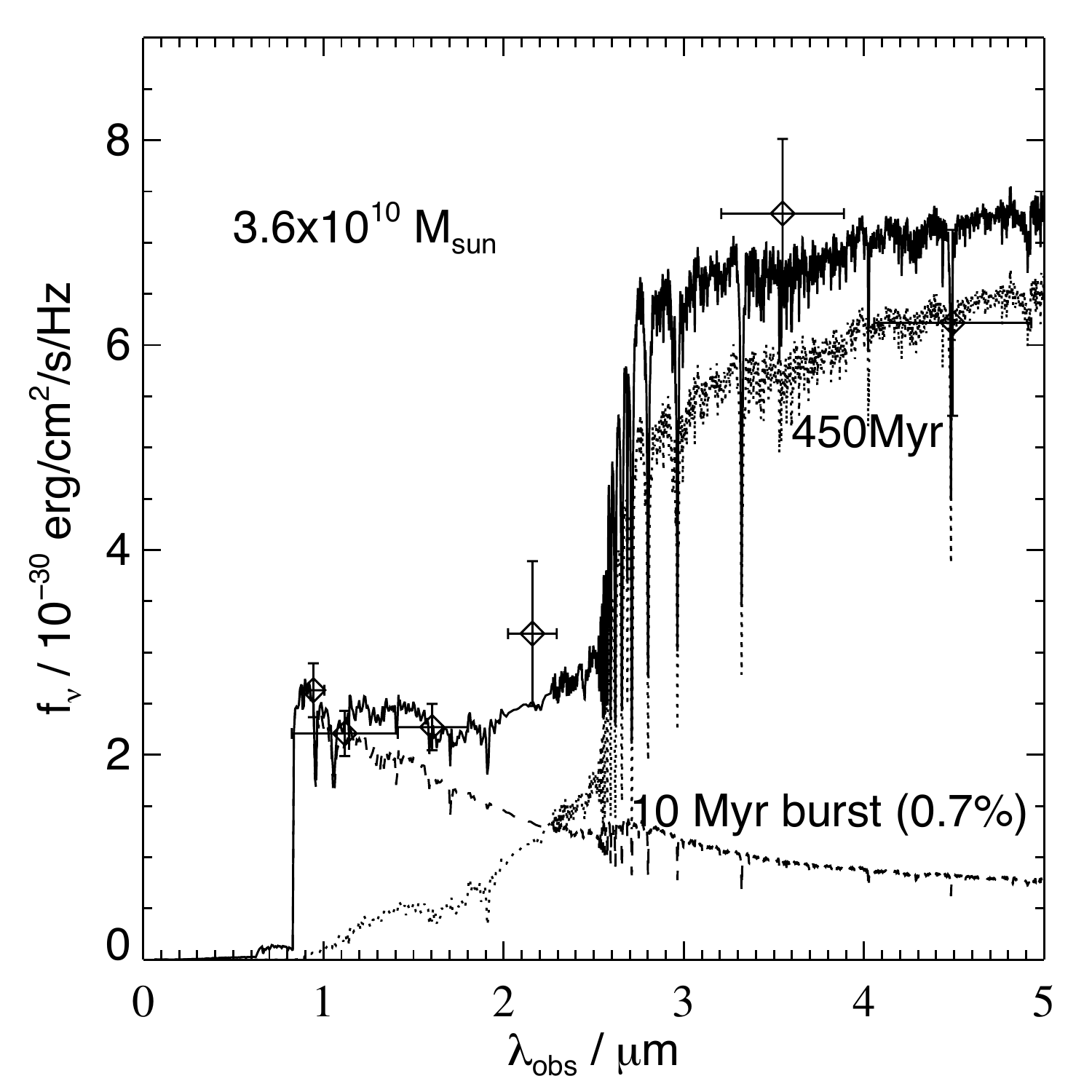}
 \includegraphics[width=3.3in]{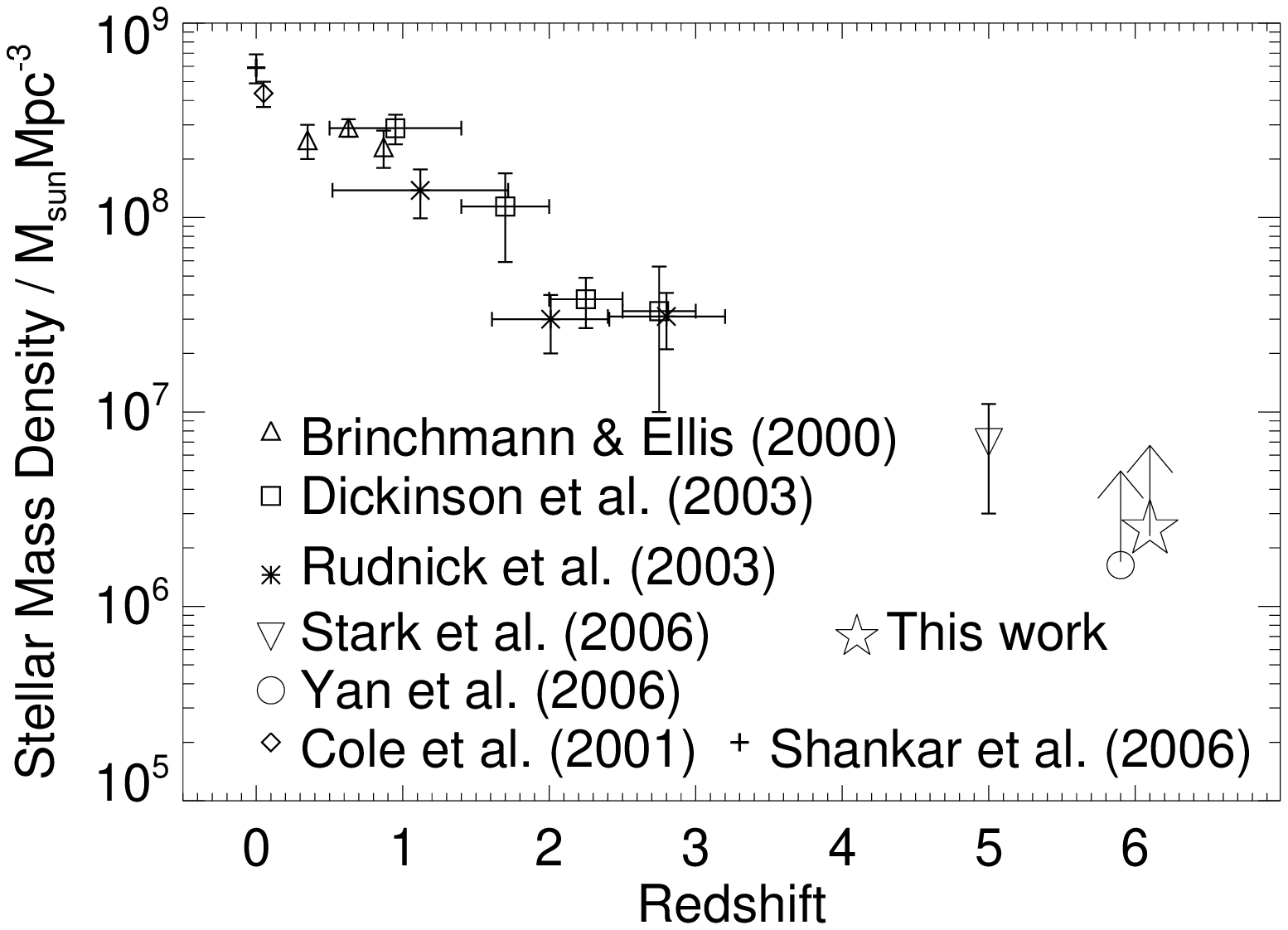} 
 \caption{{\bf Left:} the spectral energy distribution from {\em Spitzer/IRAC} and {\em HST} data for a $z\sim 6$ star-forming galaxy (Eyles et al.\ 2005).
 If we fit two stellar populations (ongoing star formation and an older burst), we find that there are stars as old as 450Myr at $z\approx 6$
 (within 900Myr of the Big Bang). The pushes the formation epoch back to $z'sim 10$, around the time to reionization.\newline
 {\bf Right:} The assembly of stellar mass at high redshift, derived from {\em Spitzer} measurements of Lyman break galaxies (Eyles et al.\ 2007). Our high redshift points are lower limits because  the Lyman break selection only picks out actively star-forming galaxies (and not post-burst systems).}
   \label{fig:spitzer}
\end{figure}

Until recently, extending this work to the $z \approx 7$ universe has been stunted by
small survey areas (from space) and low sensitivity (from the ground).
There is evidence of old stellar populations in $z\sim 4-6$  galaxies from
measurements of the Balmer break in Spitzer/IRAC imaging 
(Eyles et al.\ 2005,2007; Stark et al.\ 2007, 2009), which must have formed at higher redshift (see Fig.~\ref{fig:spitzer}).
However, the age-dating and mass-determination of these stellar populations has many uncertainties,
so searching directly for star formation at redshifts $z\ge 7$ is critical to measure the evolution of the star formation rate density, and address the role of galaxies in reionizing the universe.
 
\begin{figure}[b]
\begin{minipage}{3in}
 \includegraphics[width=3in]{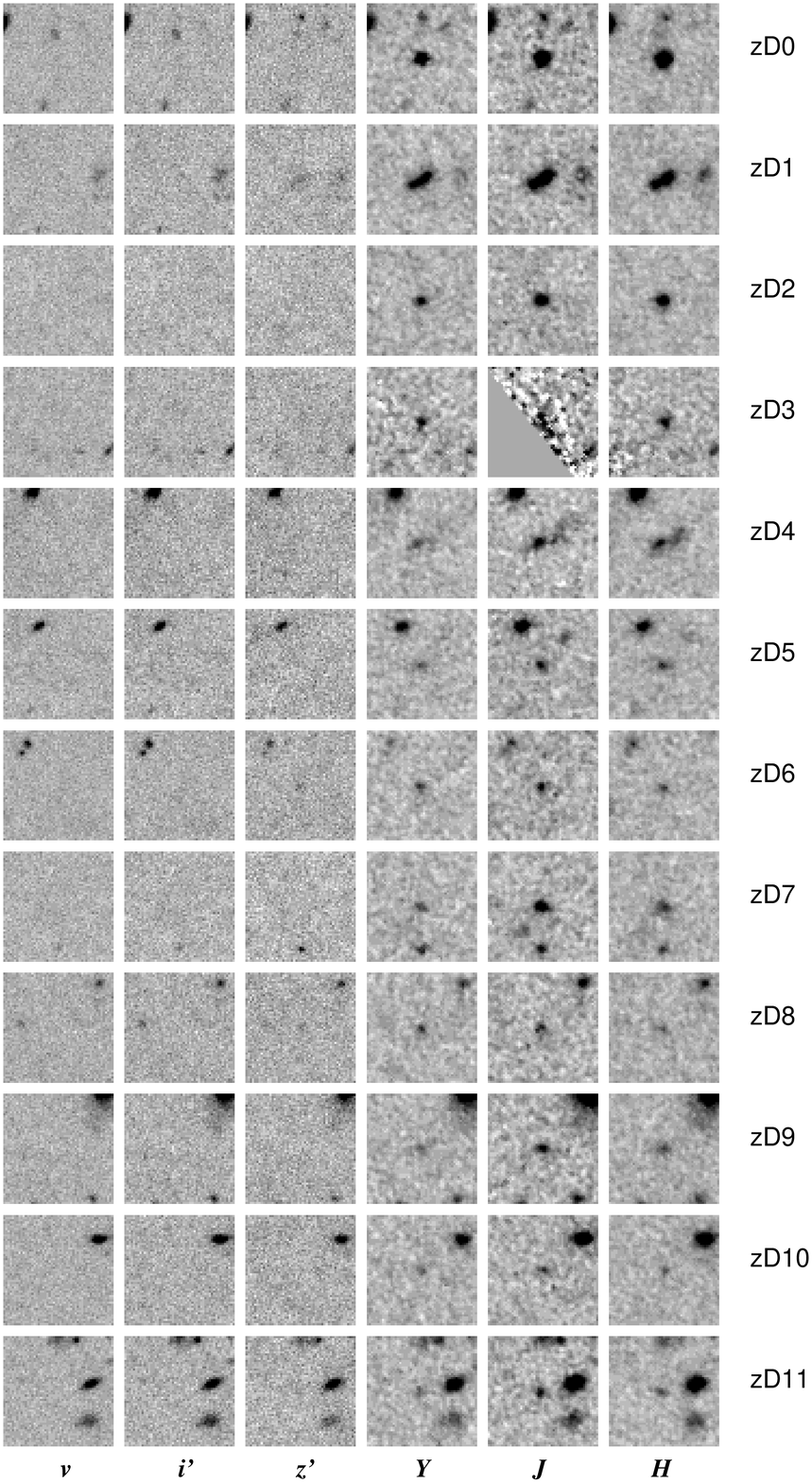} 
\end{minipage}
\begin{minipage}{2.5in}
 \includegraphics[width=2.5in]{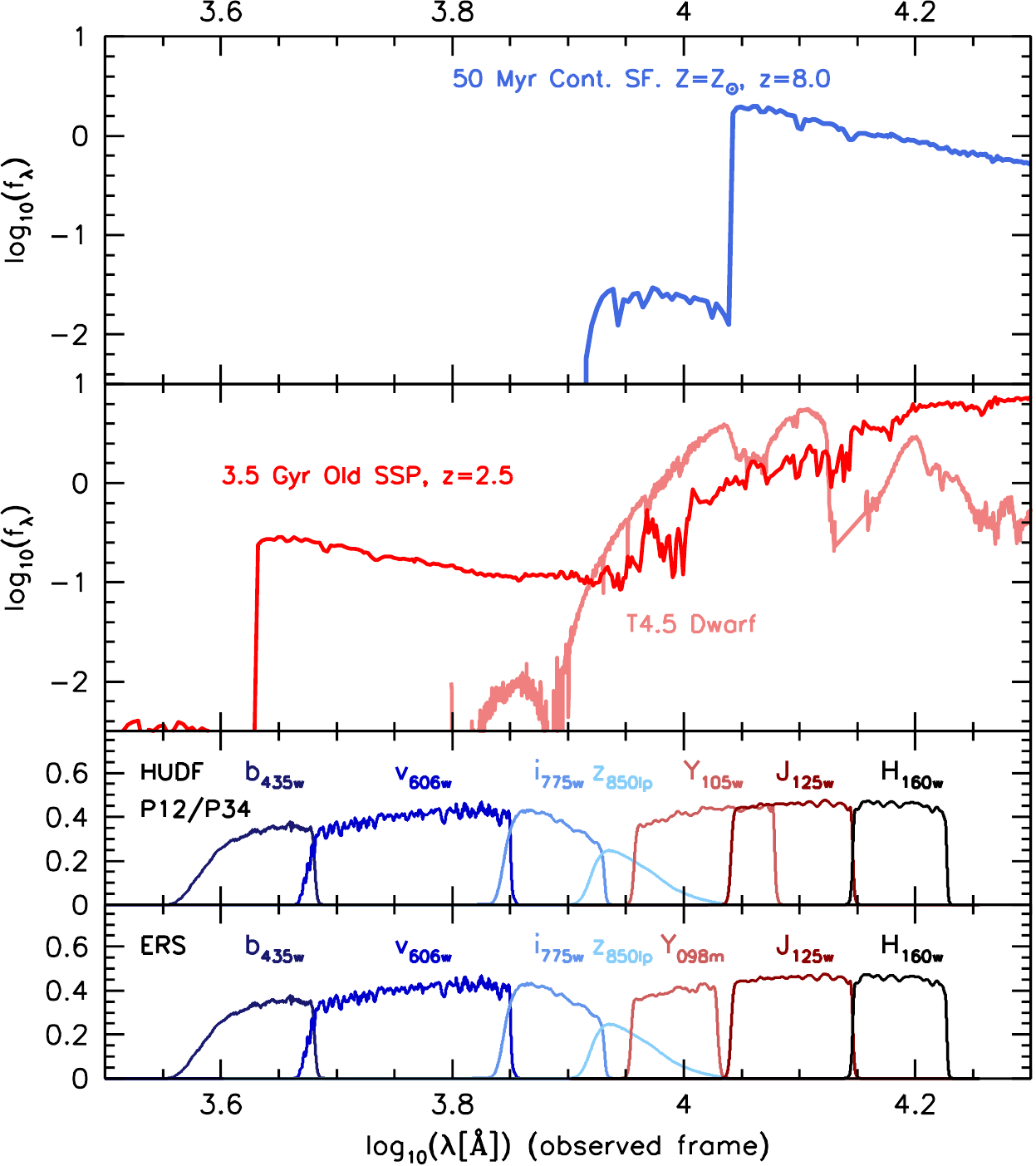} 
\includegraphics[width=2.5in]{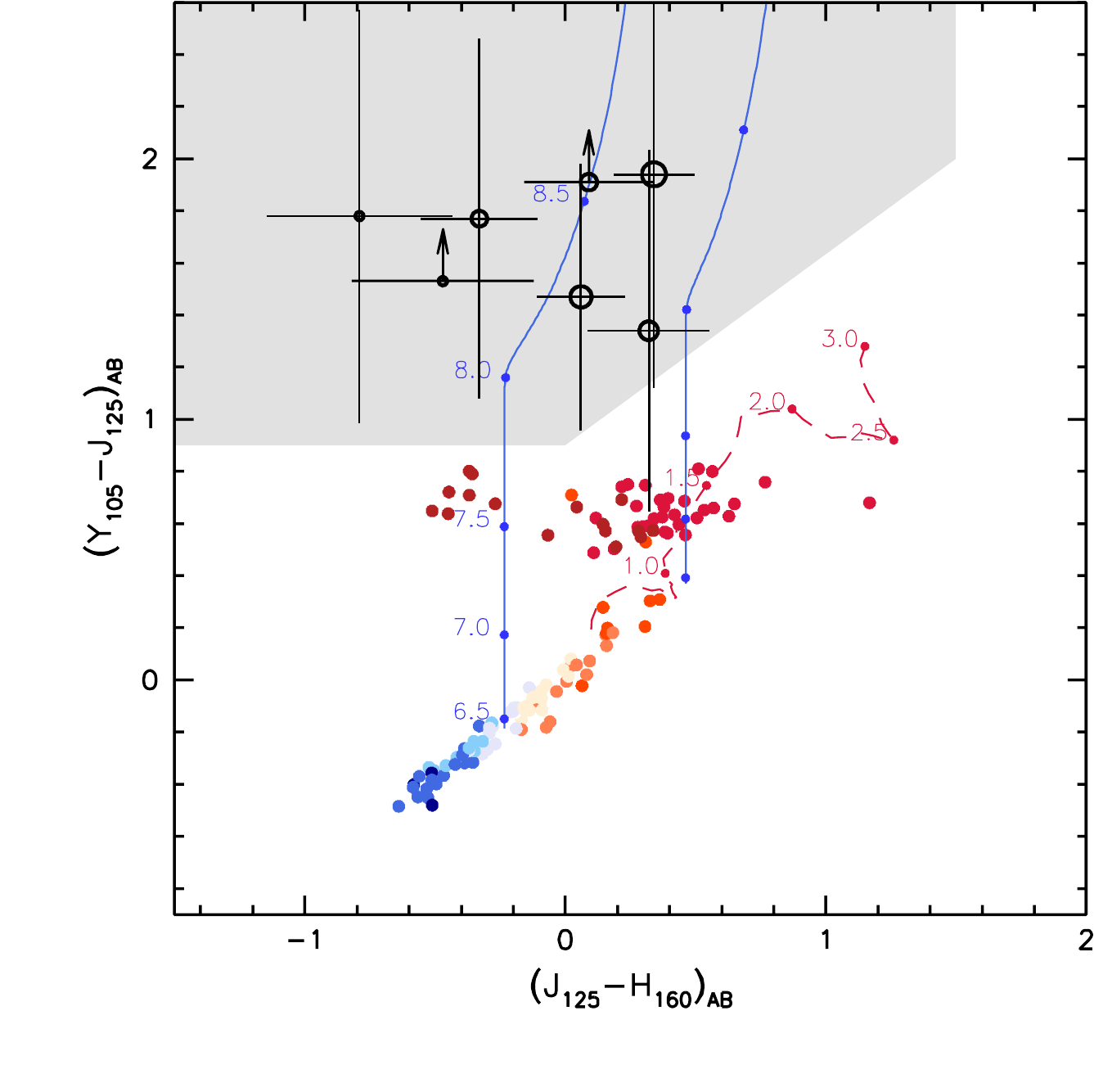}
\end{minipage}
 \caption{{\bf Left:} The $z'$-band dropouts at $z\sim 7$ identified in the WFC\,3 and ACS imaging of the {\em Hubble} Ultra Deep Field (Bunker et al.\ 2010).
 The top candidate, zD0, is actually an intermediate-redshift supernova (rather than a high redshift Lyman break galaxy) which occurred during the more recent WFC\,3 imaging from 2009 but
 was absent in the 2003 ACS images. \newline
 {\bf Top Right:} The WFC\,3 and ACS filter set on {\em HST} used, along with the spectral template for a $z\sim 7$ star-forming galaxy, and two potential interloper populations: a low mass star in our own Galaxy, and an evolved galaxy at $z\approx 2.5$ (from Wilkins et al.\ 2011a).\newline
 {\bf Bottom Right:} The use of a two-colour diagram to reject the low-redshift interlopers and select candidate $z\sim 8$ $Y$-drop galaxies (Lorenzoni et al.\ 2011). The dotted line is the locus of evolved galaxy colours as a function of redshift; the lower points are stellar colours at zero redshift, and the near-vertical lines denote star-forming galaxies with spectral slopes $\beta=-3$ (left) and $\beta=0$ (right). The points with error bars are candidate high redshift galaxies in the HUDF09 field from Lorenzoni et al.\ (2011).}
   \label{fig:WFC3}
\end{figure}

 The large field of view and enhanced
sensitivity of the recently-installed Wide Field Camera 3 (WFC3) on {\em HST}
has made great progress in identifying larger samples of $z>6$ Lyman break galaxies, 
as it covers an area 6.5 times that of the previous-generation NICMOS NIC3 camera,
and has better spatial sampling, better sensitivity and a filter set better tuned
to identifying high-redshift candidates through their colours. In Bunker et al.\ (2010) we presented an analysis of the recently-obtained WFC3 near-infrared images of the {\em Hubble} Ultra Deep Field (UDF).
We have previously used the $i'$-band and $z'$-band ACS images to identify LBGs at $z\approx 6$ through
their large $i'-z'$ colours (the $i'$-drops, Bunker et al.\ 2004). Using this ACS $z'$-band image 
in conjunction with the new WFC3 $Y$-band we could search for galaxies at $z\approx 7$, with
a spectral break between these two filters (the $z'$-drops). We also anaylsed the  $J$ and $H$-band WFC3 images to eliminate low-redshift contaminants of the $z'$-drop selection through their near-infrared colours (Fig.~\ref{fig:WFC3}), and also to look for Lyman-break galaxies at higher redshifts (the $Y$- and $J$-drops at $z\approx 8$ \& $z\approx 10$).

\begin{figure}[b]
\begin{minipage}{2.9in}
 \includegraphics[width=2.9in]{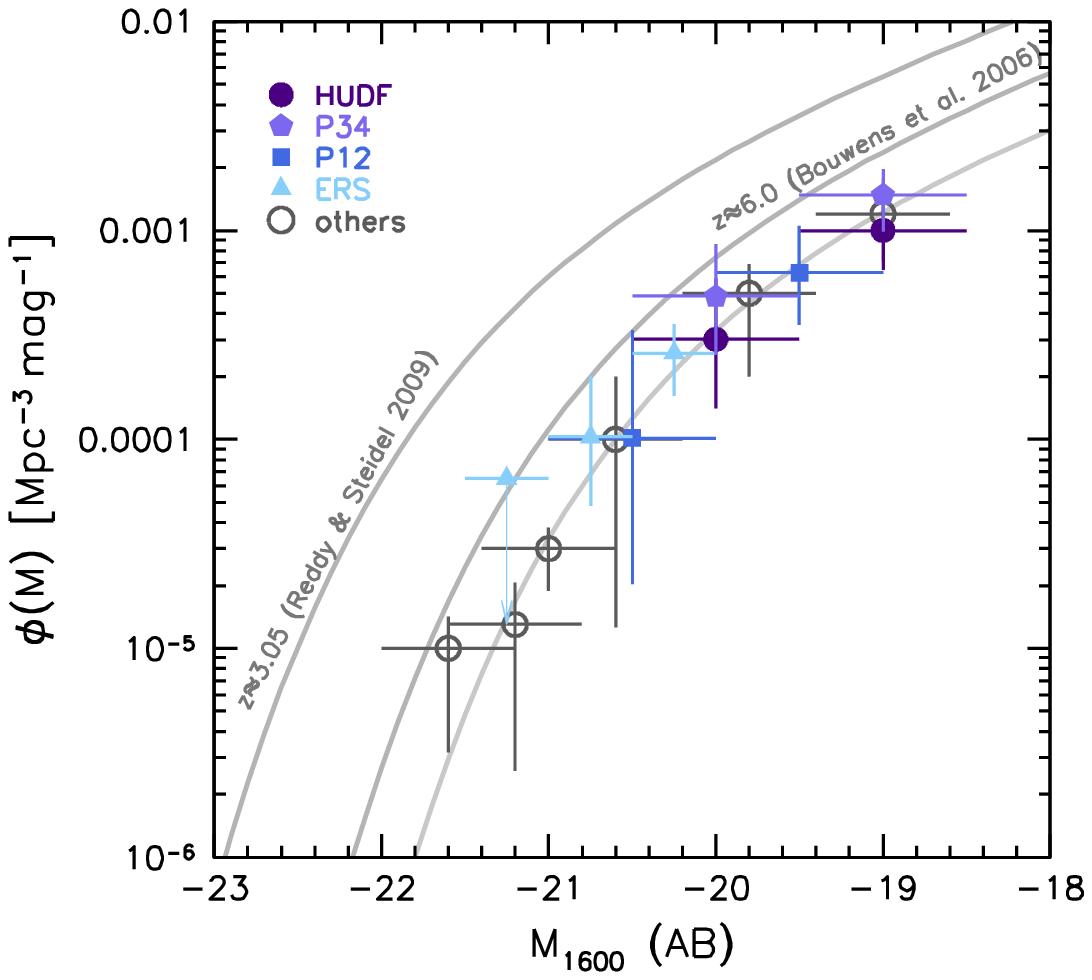} 
\end{minipage}
\begin{minipage}{2.6in}
 \includegraphics[width=2.5in]{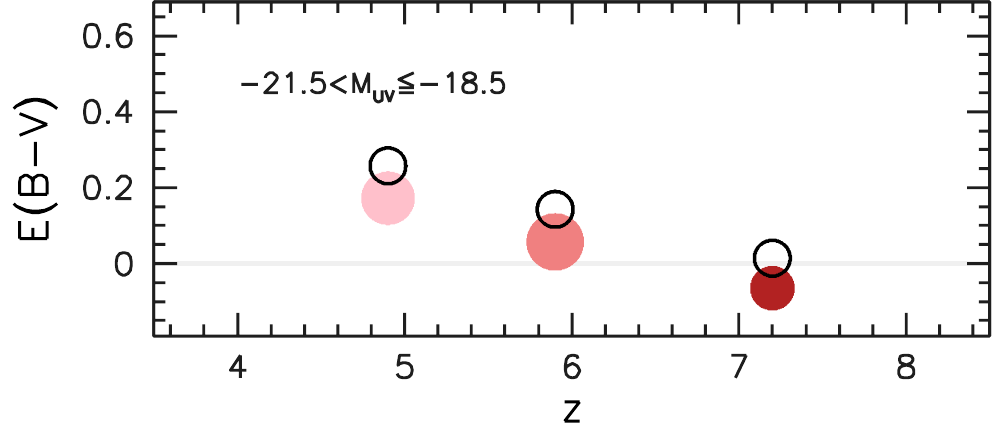} 
\includegraphics[width=2.6in]{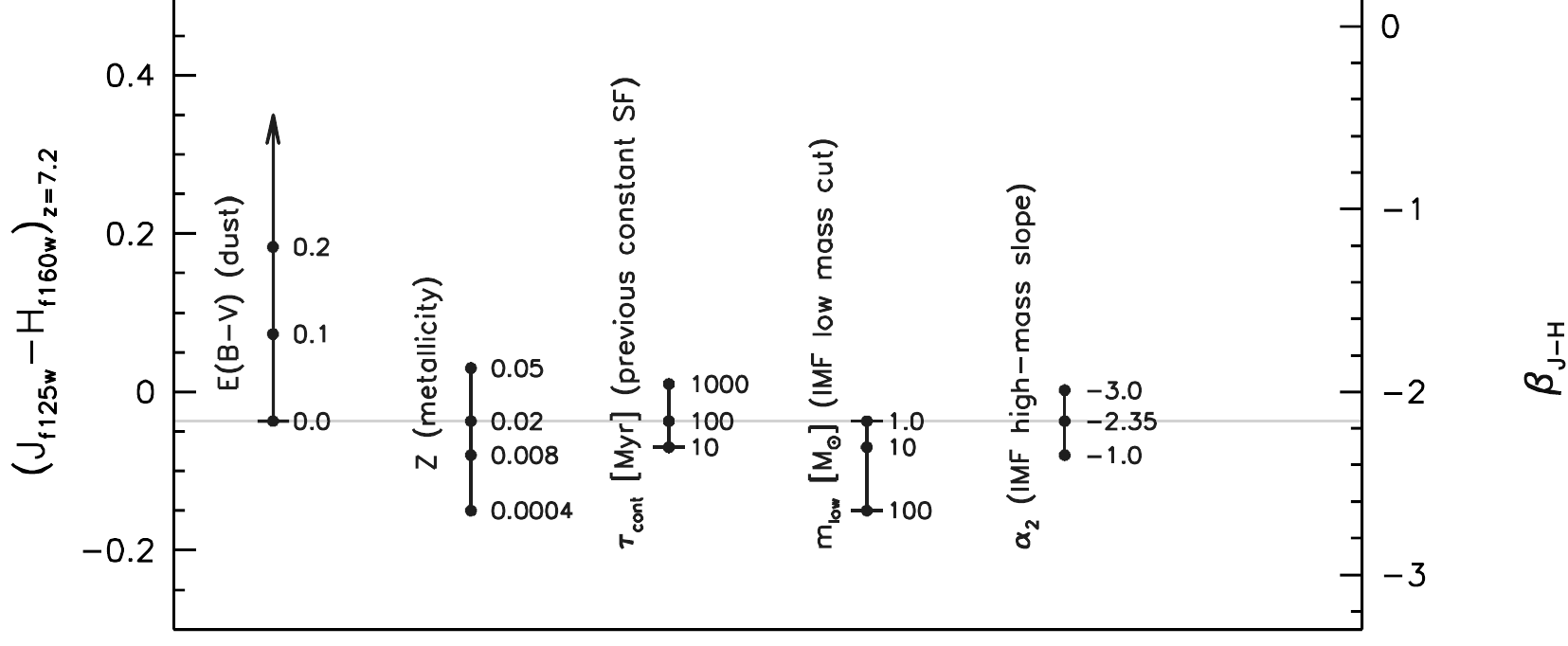}
\end{minipage}
 \caption{{\bf Left:} the rest-frame UV luminosity function of star-forming galaxies at $z\approx 7$ is dramatically lower
 than at $z\approx 2$, and even than $z\approx 6$ (from Wilkins et al.\ 2011a).\newline
 {\bf Top Right:} The evolution in the rest-UV colours of bright Lyman break galaxies; at higher redshifts, the galaxies
 are much more blue (from Wilkins et al.\ 2011b). \newline
 {\bf Bottom Right:} Various factors can affect the rest-UV spectral slopes of star-forming galaxies, such as IMF,
 metallicity, star formation history and dust (from Wilkins et al.\ 2011b). The blue colours at high redshift might imply these early galaxies
 were much less dust reddened than the corresponding lower-redshift populations.}
   \label{fig:LF}
\end{figure}

\begin{figure}[b]
\includegraphics[width=2.9in]{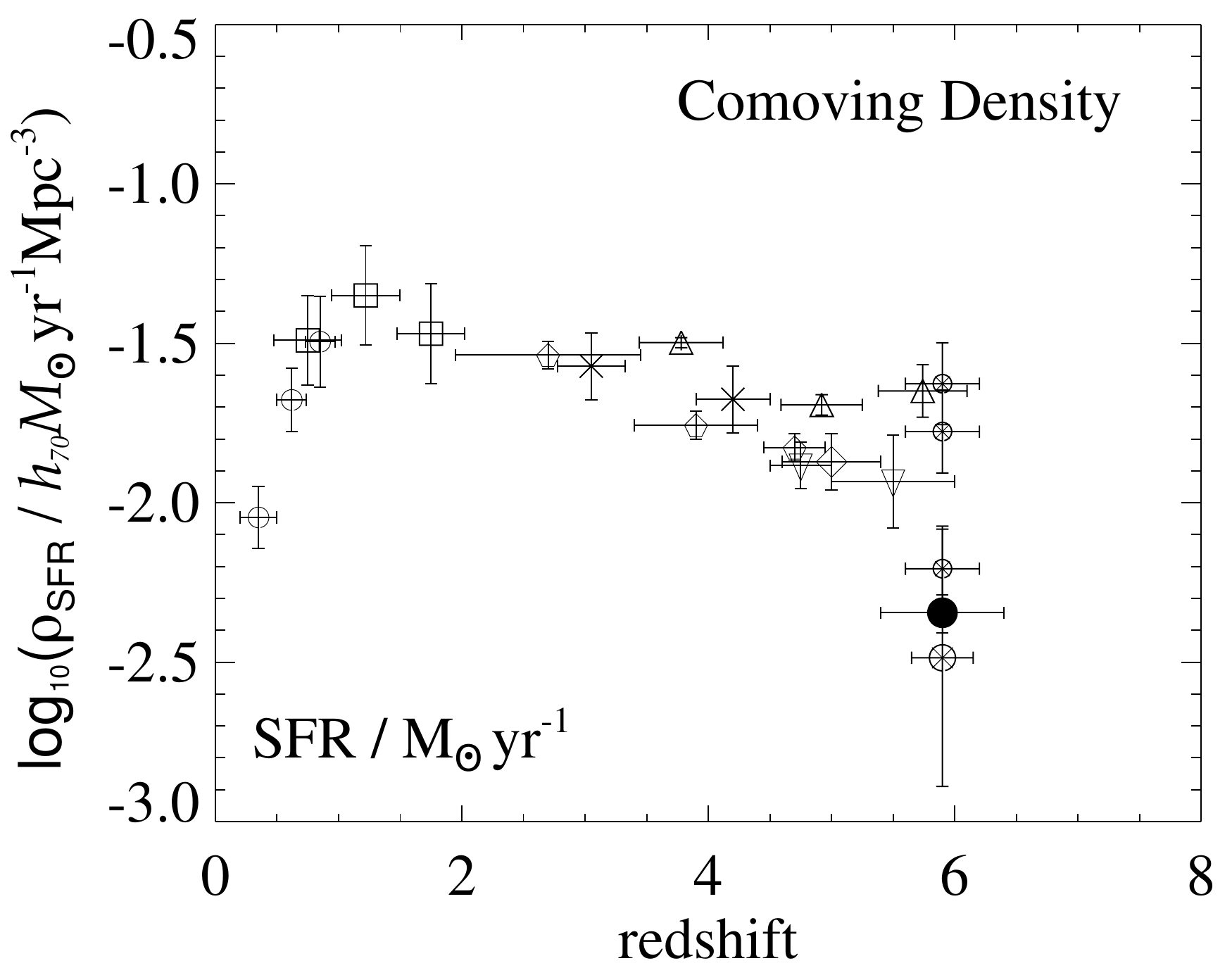}
 \includegraphics[width=2.9in]{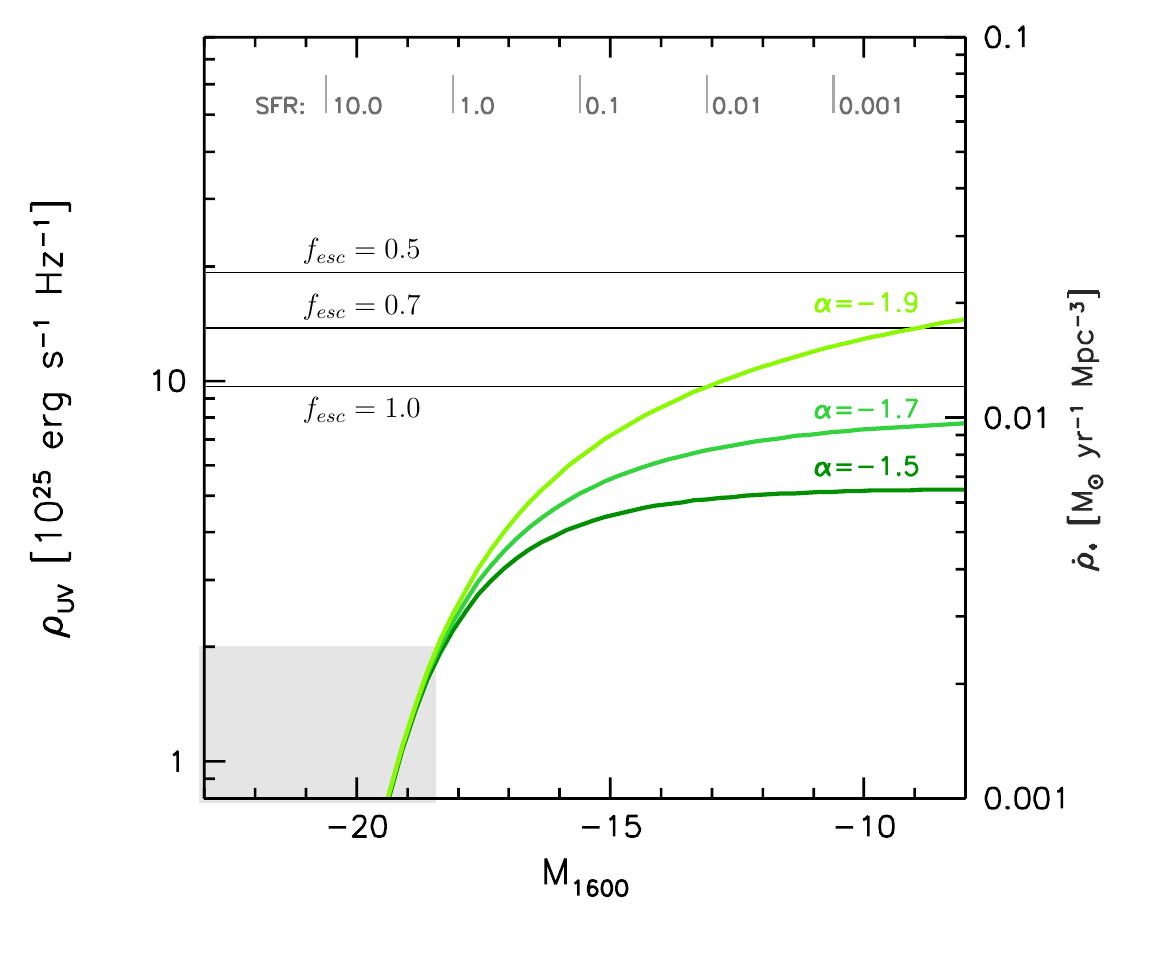} 
 \caption{{\bf Left:} the evolution of the star formation rate density (the ``Madau--Lilly" diagram) based on Lyman break galaxies. Our work
 at $z\sim 6$ (Bunker et al.\ 2004) indicates a much lower level of activity than at $z\approx 2-4$.\newline
 {\bf Right:} the ionising flux produced ($y$-axis) by the observed luminosity function of star-forming galaxies at $z\approx 8$, integrated down to a faint
 limiting magnitude (plotted on the $x$-axis). Where the integrated Schechter function curves cross the horizontal lines, there is sufficient ionising photons
 to keep the Universe ionised. As can be seen, a high escape fraction is required, and most of the photons would have to come from faint galaxies well below
 our observational limit (the shaded grey box in the lower left). Figure from Lorenzoni et al.\ (2011).}
   \label{fig:SFRD}
\end{figure}

\begin{figure}[b]
\includegraphics[width=3.1in]{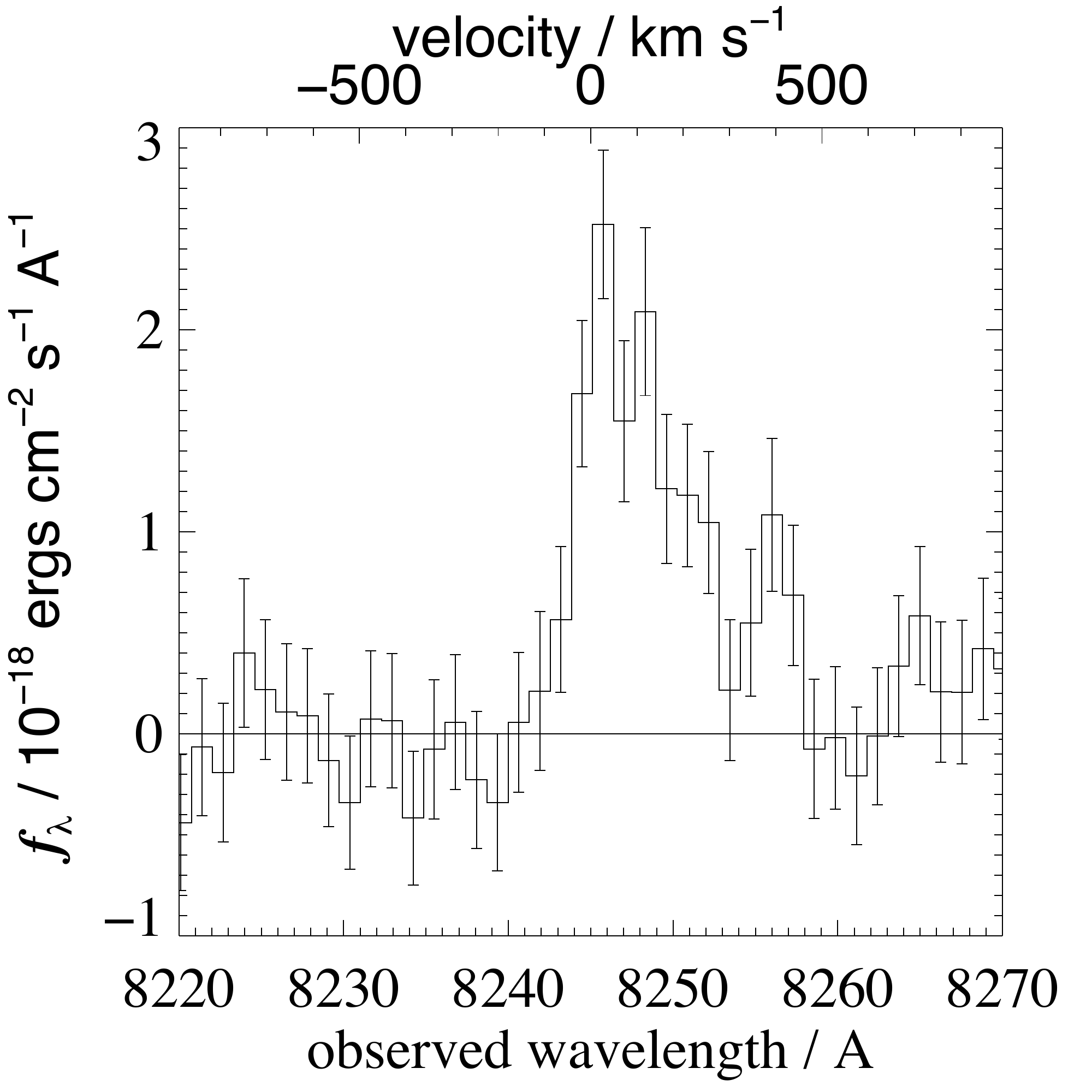}
 \includegraphics[width=2.8in]{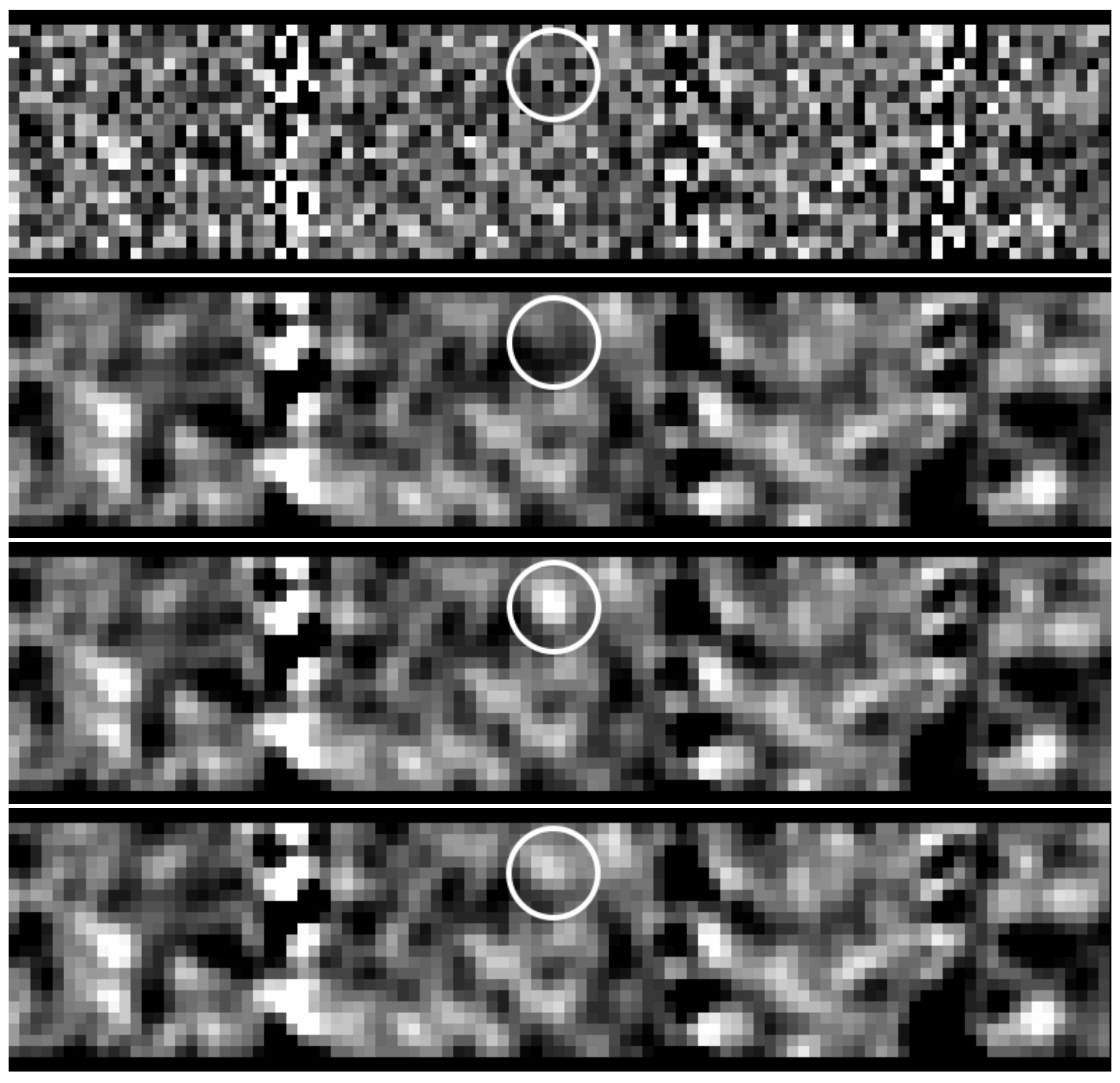} 
 \caption{{\bf Left:} A growing subset of the $z\approx 6$ Lyman break population have been spectroscopically confirmed
 through Lyman-$\alpha$ emission (figure from Bunker et al.\ 2003). At higher redshifts, $z>7$, the spectroscopy
 has been less successful, perhaps because the resonant Lyman-$\alpha$ line is absorbed by optically-thick neutral
 hydrogen in the Inter-Galactic Medium during the Gunn-Peterson era. \newline {\bf Right:} we have been unable to
 confirm the claimed spectroscopic detection of Lyman-$\alpha$
 at $z=8.55$ by Lehnert et al.\ (2010) with SINFONI in the HUDF $Y$-band drop-out $YD3$ from Bunker et al.\ (2010).
 Our  5-hour XSHOOTER spectrum is shown at the top,  and Gaussian-smoothed below. The expected position of the line
 emission is denoted by a circle. The bottom two panels have fake sources of the same flux as the Lehnert et al.\ emission line
 added in, with two different velocity widths. We should have detected the line at $\approx 4\,\sigma$ level if it was real (from
 Bunker et al.\ 2012).}
   \label{fig:spectra}
\end{figure}

\section{The Role of Star Forming Galaxies in Reionization}

We can use the observed $Y$-band magnitudes of objects in the $z'$-drop sample to estimate their star formation rate from the rest-frame UV luminosity density. 
In the absence of dust
obscuration, the relation between the flux density in the rest-UV
around $\approx 1500$\,\AA\ and the star formation rate (${\mathrm SFR}$
in $M_{\odot}\,{\mathrm yr}^{-1}$) is given by $L_{\mathrm UV}=8\times 10^{27}
{\mathrm SFR}\,{\mathrm ergs\,s^{-1}\,Hz^{-1}}$ from Madau, Pozzetti \&
Dickinson (1998) for a Salpeter (1955) stellar initial mass function
(IMF) with $0.1\,M_{\odot}<M^{*}<125\,M_{\odot}$. This is comparable
to the relation derived from the models of Leitherer \& Heckman (1995)
and Kennicutt (1998).  However, if a Scalo (1986) IMF is used, the
inferred star formation rates will be a factor of $\approx 2.5$ higher
for a similar mass range.
The redshift range is not surveyed with uniform sensitivity to UV luminosity (and hence star formation rate). We have a strong luminosity bias towards the lower end of the redshift range, due to the effects of increasing luminosity distance with redshift and also the Lyman-$\alpha$ break obscuring an increasing fraction of the filter bandwidth. Using an ``effective volume" approach (e.g., Steidel et al.\ 1999) accounts for this luminosity bias, we  infer a total star formation rate density at $z\approx 7$ of $0.0034\,M_{\odot}\,{\mathrm yr}^{-1}\,{\mathrm Mpc}^{-3}$, integrating down to $0.2\,L^*_{UV}$ of the $L^*$ value at $z=3$ (i.e.\ $M_{UV}=-19.35$)  and assuming a UV spectral slope of $\beta=-2.0$.  If instead we integrate down to $0.1\,L^*_{z=3}$ ($0.2\,L^*_{z=6}$) and the total star formation rate density is $0.004\,M_{\odot}\,{\mathrm yr}^{-1}\,{\mathrm Mpc}^{-3}$. These star formation rate densities are a factor of $\sim 10$ {\em lower} than at $z\sim 3-4$, and even a factor of $1.5-3$ below that at $z\approx 6$ (Bunker et al.\ 2004; Bouwens et al.\ 2006).

 We can compare our measured UV luminosity density at $z\approx 7$ (quoted above as a corresponding star formation rate) with that required to ionize the Universe at this redshift.
 Madau, Haardt \& Rees (1999) give the density of star formation
required for reionization (assuming the same Salpeter IMF as used in this paper):
\[
 {\dot{\rho}}_{\mathrm SFR}\approx \frac{0.005\,M_{\odot}\,{\mathrm yr}^{-1}\,{\mathrm Mpc}^{
-3}}{f_{\mathrm esc}}\,\left( \frac{1+z}{8}\right) ^{3}\,\left( \frac{\Omega_{b}\,h^
2_{70}}{0.0457}\right) ^{2}\,\left( \frac{C}{5}\right)
 \]
We have updated equation 27 of Madau, Haardt \& Rees (1999) for a more recent concordance cosmology estimate of the baryon
density from Spergel et al.\ (2003).
$C$ is the clumping factor of neutral
hydrogen, $C=\left< \rho^{2}_{\mathrm HI}\right> \left< \rho_{\mathrm
    HI}\right> ^{-2}$. Early simulations suggested $C\approx 30$ (Gnedin \&
Ostriker 1997), but more recent work including the effects of reheating implies a lower
concentration factor of  $C\approx 5$ (Pawlik et al.\ 2009). 
The escape fraction of ionizing photons ($f_{\mathrm esc}$) for
high-redshift galaxies is highly uncertain (e.g., Steidel, Pettini \&
Adelberger 2001, Shapley et al.\ 2006), but even if we take $f_{\mathrm esc}=1$ (no absorption by
H{\scriptsize~I}) and a very low clumping factor, this estimate of the star formation density required
is a factor of 1.5--2 higher than our measured star formation density at
$z\approx 7$ from $z'$-drop galaxies in the UDF.  For faint end slopes of $\alpha ~-1.8\rightarrow-1.6$
galaxies with $L>0.2\,L^{*}$ account for $24-44$\% of the total
luminosity (if there is no low-luminosity cut-off for the Schechter function), so even with a steep faint-end slope  at $z\approx 7$ we still fall short of the required
density of Lyman continuum photons required to reionize the Universe, unless the escape fraction is implausibly high ($f_{\mathrm esc}>0.5$) and/or the faint end slope is $\alpha<-1.9$ (much steeper than observed at $z=0-6$). At $z\approx 8$, the situation is even more extreme (Lorenzoni et al.\ 2011), as the luminosity function is even lower (Fig.~\ref{fig:SFRD}).

However, the assumption of a solar metallicity Salpeter IMF may be flawed: the colours of $z\sim 6$ $i'$-band drop-outs are very blue (Stanway, McMahon \& Bunker 2005), with $\beta<-2$, and the new WFC3 $J$- and $H$-band images show that the $z\approx 7$ $z'$-drops also have blue colours on average (Figure~\ref{fig:LF}). A slope of $\beta<-2$ is bluer that for continuous star formation with a Salpeter IMF, even if there is no dust reddening. Such blue slopes could be explained through low metallicity,
 or a top-heavy IMF, which can produce between 3 and 10 times as many ionizing photons for the same 
 1500\,\AA\ UV luminosity (Schaerer 2003 -- see also Stiavelli, Fall \& Panagia 2004). Alternatively, we may be seeing galaxies at the onset of star formation, or with a rising star formation rate (Verma et al.\ 2007), which would also lead us to underestimate the true star formation rate from the rest-UV luminosity. We explore the implications of the blue UV spectral slopes in $z\ge 6$ galaxies in Wilkins et al.\ (2011b).
  
We have also targeted many of our high redshift Lyman break galaxies with spectroscopy from the ground, using Keck/DEIMOS (Bunker et al.\ 2003), Gemini/GMOS (Stanway et al.\ 2004) and most recently VLT/XSHOOTER (Bunker et al.\ 2012, Fig.~\ref{fig:spectra}). While we have seen Lyman-$\alpha$ emission is several $z\approx 6$ galaxies, the frequency seems to drop at $z>7$, perhaps due to absorption of this line in the increasingly-neutral IGM.

\section{Conclusions}

The availability of WFC3 and ACS on the {\em Hubble} Space Telescope has made the discovery of Lyman break galaxies at $z>5$ possible.
From these galaxies, we have shown that the star formation rate density declines at $z>5$.
Based on their rest-frame UV luminosity function, it appears that a high escape fraction for Lyman continuum photons and a huge
contribution from faint sources well below current detection limits would be needed for star forming galaxies to keep the Universe ionised
at these redshifts. We have been able to measure stellar masses and ages from Spitzer in some instances. We also have discovered that
the high redshift Lyman break galaxies have much bluer rest-frame UV colours than the $z\approx 4$ population, perhaps indicative
of less dust obscuration at higher redshifts. Spectroscopic follow-up is ongoing, but we have confirmed Lyman-$\alpha$ emission
from several $z\approx 6$ galaxies, although the success rate drops at $z>7$, perhaps due to the onset of Gunn-Peterson absorption.

\section*{Acknowledgments}

I gratefully acknowledge the involvement of my collaborators on these projects -- Stephen Wilkins, Elizabeth Stanway, Richard Ellis, Silvio Lorenzoni, Joseph Caruana, Matt Jarvis, Samantha Hickey, Daniel Stark, Mark Lacy, Kuenley Chiu, Richard McMahon and Laurence Eyles. I thank the IAU and the conference organisers for an excellent Symposium.




\end{document}